\documentclass[aps,twocolumn,showpacs]{revtex4}

\usepackage{graphicx}

\newcommand{\be}{\begin{equation}}
\newcommand{\ee}{\end{equation}}
\newcommand{\ba}{\begin{eqnarray}}
\newcommand{\ea}{\end{eqnarray}}
\def\diff{\mathop{\rm\mathstrut d\!}\nolimits}

\begin{document}

\title{Smectic ordering of parallel hard spherocylinders:  \\
An entropy-based Monte Carlo study}
\author{
D.~Costa$^\dag$, F. Saija$^\ddag$, and
P.~V.~Giaquinta$^\dag$\footnote{Corresponding author; email:
{\tt Paolo.Giaquinta@unime.it}}} 
\affiliation{\dag~Istituto Nazionale per la Fisica della Materia (INFM) and
Universit\`a degli Studi di Messina, Dipartimento di Fisica
Contrada Papardo, C.P. 50 -- 98166 Messina, Italy \\
\ddag~CNR -- Istituto per i Processi Chimico-Fisici, sez. Messina
Via La Farina 237 --  98123 Messina, Italy}

\begin{abstract}
We investigated the nematic to smectic transition undergone by
parallel hard spherocylinders in the framework provided by the
residual multi-particle entropy (RMPE) formalism. The RMPE is
defined as the sum of all contributions to the configurational
entropy of the fluid which arise from density correlations
involving more than two particles. The vanishing of the RMPE
signals the structural changes which take place in the system for
increasing pressures. Monte Carlo simulations carried out for
parallel hard spherocylinders show that such a one-phase ordering
criterion accurately predicts also the nematic-smectic transition
threshold notwithstanding the almost continuous character of the
transition. A similar quantitative correspondence had been already
noted in the case of an isotropic fluid of freely rotating hard
spherocylinders undergoing a transition to a nematic, smectic or
solid phase. The present analysis confirms the flexibility of the
RMPE approach as a practical and reliable tool for detecting the
formation of mesophases in model liquid-crystal systems.
\end{abstract}

\pacs{64.60.-i, 64.70.Md, 61.20.Ja, 65.40.Gr}

\maketitle

\section{Introduction}

Recently we investigated the formation of liquid-crystalline
mesophases in a fluid of hard spherocylinders, i.e., cylindrical
segments of length $L$ and diameter $D$ capped at each end by a
hemisphere of the same diameter~\cite{micali}. More specifically,
we analyzed the ordering of the homogeneous and isotropic fluid
into a nematic, smectic or solid phase within the framework
provided by the one-phase entropy-based criterion originally
proposed by Giaquinta and Giunta~\cite{gg}. This criterion can be
implemented through the well known multi-particle correlation
expansion of the configurational entropy~\cite{ng}:
\be\label{s:sum} s_{\rm ex}=\sum_{n=2}^{\infty} s_{n} \,. \ee In
the above formula $s_{\rm ex}$ is the excess entropy per particle
in units of the Boltzmann constant and $s_{n}$ is the $n$-body
entropy that is obtained upon re-summing spatial correlations
between up to $n$ particles. In particular, the pair entropy per
particle of a homogeneous and isotropic fluid of nonspherical
molecules can be written as \cite{lapa}: 
\ba\label{eq:s2mol}
s_{2}(\rho) & = &  -\frac{1}{2} \frac{\rho}{\Omega^2} \int \{ g ({\bf
r},\omega^2) \ln [g({\bf r},\omega^2)] \nonumber \\
 && -   g ({\bf r},\omega^2) + 1\}
\diff {\bf r} \diff \omega^2 \,, \ea 
where $g({\bf r},\omega^2)$
is the pair distribution function (PDF) which depends on the
relative separation $\bf r$ between two molecules and on the set
of Euler angles $\omega^2\equiv[\omega_{1},\omega_{2}]$ that
specifies their absolute orientations in the laboratory reference
frame. The quantity $\Omega$ represents the integral over the
Euler angles of one molecule, while $\rho$ is the particle number
density.

In order to identify the ordering threshold of the fluid, we
monitor the behavior~---~as a function of the number density~---~of
the so-called residual multiparticle entropy (RMPE). This quantity
is defined as the difference \be \label{s:sex} \Delta s \equiv
s_{\rm ex} - s_{2} \,. \ee At variance with the pair entropy, the
RMPE exhibits a nonmonotonic behavior as a function of $\rho$. In
particular, it is negative at low densities, and becomes positive
as a more ordered phase is approached. The relevance of the
condition $\Delta s = 0$ as a one-phase ordering criterion has
been documented for a variety of phase transitions, both in
continuous fluids as well as in lattice-gas model
systems~\cite{spg1}.

Even for systems composed of non-spherical molecules, such as hard
spherocylinders, it turns out that the ordering thresholds
detected through the zero-RMPE condition systematically correlate
with the corresponding phase-transition points, whatever the
nature of the higher-density phase coexisting with the isotropic
fluid~\cite{micali}. Cuetos and coworkers have successfully
applied the RMPE criterion to hard spherocylinders with an
attractive square well and to spherocylinders with a soft
repulsive core~\cite{cuetos}.

In this paper we intend to analyze the predictions of the RMPE
approach in the case of the nematic-smectic transition undergone
by parallel hard spherocylinders with aspect ratio $L/D=5$. The
onset of smectic order out of a nematic phase represents the next
step in the process which, upon compression of the isotropic
fluid, eventually leads to the formation of the fully crystalline
solid. The phase behavior of this model has been investigated with
several numerical simulations~\cite{slf,verfre,koda} as well as
theoretical studies~(see e.g.~\cite{hol,tay,baus} and references
contained therein) over the whole $L/D$ range. Such studies have
ascertained the existence of a second-order phase transition from
a low-density nematic state to an intermediate smectic phase. On
the other hand, the general features of the higher-density region
of the phase diagram has long been debated as far as the stability
a columnar phase is concerned~\cite{verfre}.

The paper is organized as follows: In Sections~2 and~3 we
describe the theoretical framework and the numerical simulation
technique. The results are presented in Section~4 while Section~5
is devoted to concluding remarks.

\section{Theory}

The formalism developed in~\cite{micali} for the isotropic fluid
needs to be modified in order to describe the nematic and smectic
mesophases. Even for uniaxial molecules, the numerical computation
of the full pair distribution function $g({\bf r},\omega^2)$ is a
formidable task. Costa and coworkers have already shown that, in
order to reproduce the phase boundaries of the isotropic fluid, it
is enough to take into account the dependence of $g({\bf
r},\omega^2)$ on the centers-of-mass separation $r$ and on the
angle $\theta$ formed by the molecular axes of two
spherocylinders~\cite{micali}. This angle shows up as the critical
parameter which accounts, by itself, for the reduction of
orientational states that is ultimately offset by the gain of
translational entropy at high densities~\cite{ons}. On the other
hand, it is rather obvious that the angle $\theta$ conveys no
useful information on the structural process that may eventually
lead to the formation of a smectic phase out of a nematic phase.

In order to resolve ordering effects associated with the
orientational degrees of freedom on one side, and with the
modulation of the density along the nematic director on the other
side, we shall restrict our analysis to a system composed of
parallel spherocylinders. Assuming that the molecules are
perfectly aligned does not significantly alter the main aspects of
the phenomenology that we want to investigate. As emphasized
above, different mechanisms drive the phase transformation of an
isotropic or of a nematic fluid into a more ordered phase.
Moreover, at the nematic-smectic transition threshold the nematic
fluid is typically characterized by a very high degree of relative
alignment of the molecules. For example, the nematic order
parameter is about 0.85-0.89 for spherocylinders with aspect ratio
$L/D=5$~\cite{jawil}. The constraint on the relative orientation
of the particles hugely simplifies the expression of the PDF. In
fact, in order to specify the spatial configuration of a given
spherocylinder relative to that fixed at the origin one just needs
two parameters: the distance $r$ and the angle $\vartheta$ formed
by $\bf r$ and by the nematic director. The other polar angle
$\phi$ is actually averaged out on account of the macroscopic
cylindrical symmetry of the model. We emphasize that in this case,
at variance with the freely rotating model already investigated
in~\cite{micali}, no approximation is needed in order to compute
the PDF of the system. Correspondingly, the pair entropy of the
fluid can be written as: 
\ba\label{eq:s2} s_{2}(\rho) & = &  -\pi \rho
\int_0^\infty r^2\diff r \int_0^\pi \sin \vartheta \diff \vartheta
\nonumber \\
& &  \{ g(r,\vartheta) \ln [g(r,\vartheta)] 
- g(r,\vartheta) + 1\}
\,. \ea 
Using a formalism analogous to that introduced
in~\cite{micali}, we can extract from Eq.~(\ref{eq:s2}) the
excluded-volume contribution to the pair entropy of the fluid.
This contribution arises from the space integration carried out
over the regions where $g(r,\vartheta) = 0$: \be\label{s:s2}
s_{2}=- B_2\rho + s_{2}^{\prime} \;. \ee In Eq.~(\ref{s:s2})
${B_2}$ is the second virial coefficient of hard parallel
spherocylinders that is just four times the volume, $V_{\rm hsc}$,
of a spherocylinder: \be B_2 = 4[(\pi/4)D^2L+(\pi/6)D^3] \;. \ee
The residual contribution, $s_{2}^{\prime}$, explicitly accounts
for the decrease of the pair entropy associated with the onset of
interparticle correlations at short and medium-range distances.
For parallel spherocylinders, a further separation of this term
into a translational and an orientational contribution as done
in~\cite{micali} does not make much sense.

The total excess entropy of the nematic fluid can be evaluated
upon integrating the equation of state (EoS): \be\label{s:free_ex}
s_{\rm ex}(\rho) = s_{\rm ex} (\bar{\rho}) -
\int_{\bar{\rho}}^{\rho} \left[ \frac{{\beta} P} {\rho'} - 1
\right] \frac{\diff{\rho'}}{\rho'} \,. \ee In
Eq.~(\ref{s:free_ex}) $P$ is the pressure, $\beta$ is the inverse
temperature in units of the Boltzmann constant, and $\bar{\rho}$
is the density of a suitably chosen reference state. The EoS of
the fluid was obtained through a Monte Carlo sampling of the
system at different pressures, while the quantity $s_{\rm
ex}(\bar{\rho})$ was calculated using the Widom test-particle
insertion method~\cite{widom,smfr} in order to calculate the
excess chemical potential, $\mu_{\rm ex}$, at low enough densities.
The excess entropy can then be obtained through the thermodynamic
equation: \be\label{eq:muex} s_{\rm ex} = - \beta \mu_{\rm ex} +
\frac{\beta P}{\rho} -1 \,. \ee

\section{Simulation}

We investigated the phase diagram of a system of parallel hard
spherocylinders with elongation $L/D = 5$, spanning a density
range which goes from a rather dilute nematic state up to the
smectic transition threshold. We carried out Monte Carlo (MC)
simulations at constant pressure as is usual for systems of
non-spherical hard-core particles where it may be difficult to
calculate the equation of state in a constant-volume simulation
through the contact values of the distribution
functions\cite{smfr}.

The typical sample was composed of $N=1500$ particles, aligned
along the $z$ axis and enclosed in an orthorhombic box with edges
$L_x = L_y = 1/3 L_z$. In order to quantify the influence of the
size of the system as well as of the shape of the simulation box,
we also investigated the behavior of a system composed of $500$
particles aligned along the main diagonal of a cubic box and of a
fluid of $768$ particles enclosed in a orthorhombic cell with the
same relative dimensions used for the main sample.

All thermodynamic states at constant pressure were sequentially
generated from a translationally disordered low-density
configuration upon gradually compressing the nematic fluid. The
equilibration period was typically $10^{5}$ MC cycles, a cycle
consisting of an attempt to change sequentially the center-of-mass
coordinates of each molecule followed by an attempt to modify the
volume of the sample. Simulation data were obtained by generating
chains consisting of $5\times10^{5} - 20\times10^{5}$ MC cycles,
depending on the pressure. Equilibrium averages and standard
deviations were computed by dividing chains into independent
blocks. During the production runs, we cumulated different
histograms of the PDF. In particular, $g(r,\vartheta)$ was sampled
at intervals $\Delta r$ and $\Delta \vartheta$ of $0.05D$ and
$1^\circ$, respectively. Different choices of $\Delta r$ and
$\Delta \vartheta$ were investigated for $P^*=1.0$ and $P^* =
2.0$. As for the Widom insertion method, $100$ trial insertions
per MC cycle turned out to be sufficient to insure a stable
statistics for the excess chemical potential of the fluid.

In the presentation of the results, we shall refer to the reduced
density  $\rho^* = \rho/\rho_{\rm cp}$, where $\rho_{\rm cp} =
2/[\sqrt{2}+(L/D)\sqrt{3}]$ is the maximum density attained by
parallel spherocylinders at close packing, and to the reduced
pressure $P^*=\beta PV_{\rm hsc}$.

\section{Results}

\begin{figure*}[t]
\begin{center}
\includegraphics[width=9cm,angle=-90]{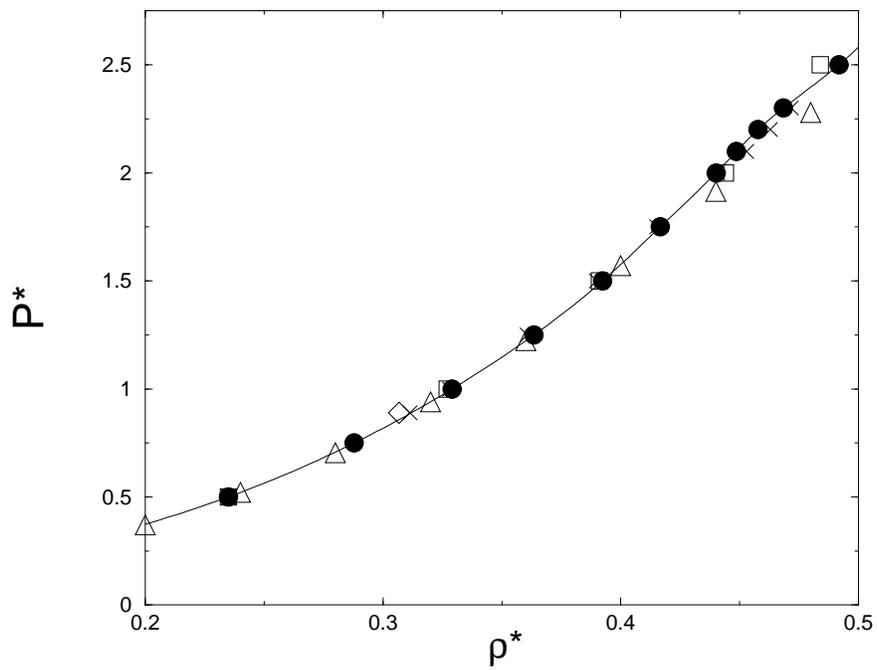}
\caption{Equation of state of parallel hard spherocylinders with
aspect ratio $L/D=5$. Line with solid circles and crosses: this
work for $N=1500$ and $N=500$ molecules, respectively. Triangles, squares
and diamond: simulations by Veerman and
Frenkel~\protect\cite{verfre}, Stroobants and
coworkers~\protect\cite{slf}, and Koda and
Ikeda~\protect\cite{koda}, respectively. The error bars are
systematically smaller than the size of the markers.
}\label{fig:1}
\end{center}
\end{figure*}

The current MC results for the EoS of the model are presented in
Fig.~\ref{fig:1}. The comparison between the data collected for
$1500$ particles (in an orthorombic box) and those for $500$
particles (in a cubic box) shows that the EoS of the fluid is
rather weakly affected by the size of the sample only at high
densities. The results for $768$ particles in an orthorombic box
were not reported in the graph since they can be hardly resolved
from those pertaining to the largest size investigated. The
present EoS is in good agreement with that computed by other
authors~\cite{slf,koda} over the whole density range explored,
apart from a modest deviation, at high densities, from the
molecular dynamics results obtained by Veerman and
Frenkel~\cite{verfre}. Both these authors as well as Stroobants
and coworkers~\cite{slf} noted a weak change of slope in the EoS
at the transition that was estimated to occur for $\rho^*=0.46$.
Other structural properties of the system were evaluated
in~\cite{slf} such as the distribution functions associated with
parallel and perpendicular correlations, 
below and above the observed cusp. Given the absence of
either translational order within the layers or hysteresis effects
associated with a compression/expansion cycle, Stroobants and
coworkers concluded that the nematic phase actually transforms
into a smectic phase through a continuous phase transition.
Veerman and Frenkel later estimated the transition density by
studying the critical slowing down of the intermediate scattering
function evaluated along the nematic
director~\cite{verfre}. Their result agrees with that reported by
Stroobants and coworkers~\protect\cite{slf}.

\begin{figure*}[t]
 \dimen0=\textwidth \advance\dimen0 by -\columnsep \divide\dimen0 by 4
 \noindent\begin{minipage}[!t]{\dimen0}
 \flushleft
 {\includegraphics[width=3.5cm,angle=-90]{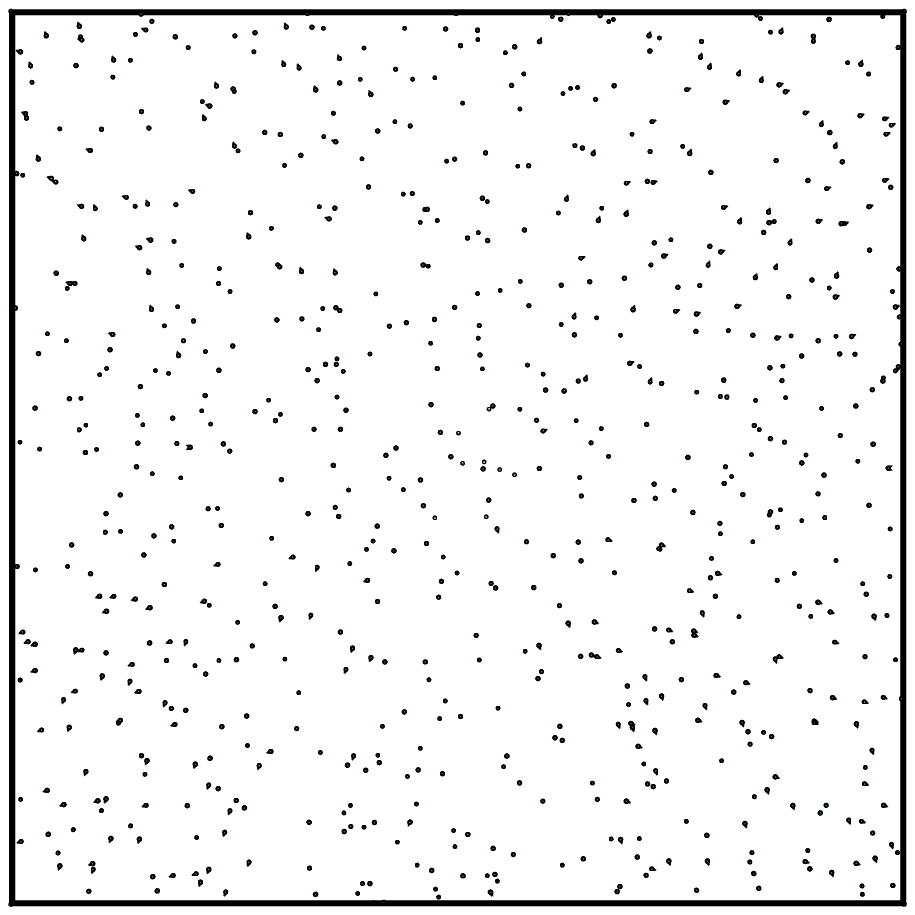}\bigskip}
 {\includegraphics[width=3.5cm,angle=-90]{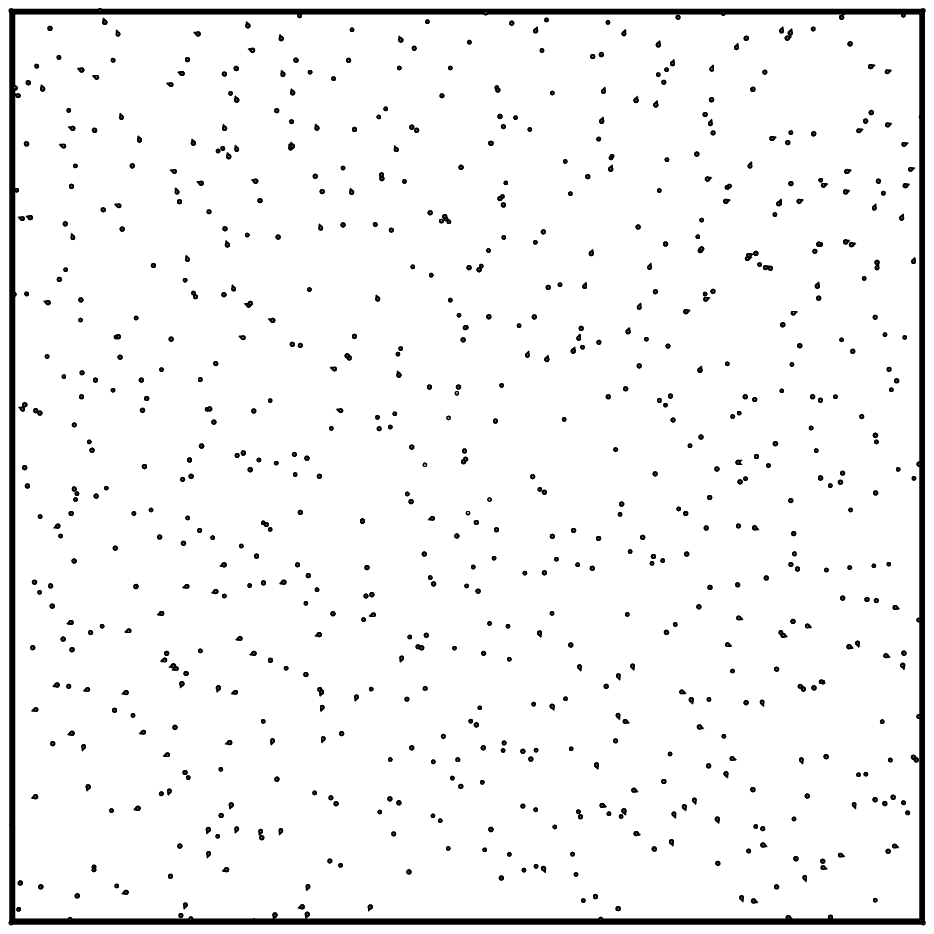}\bigskip}
 {\includegraphics[width=3.5cm,angle=-90]{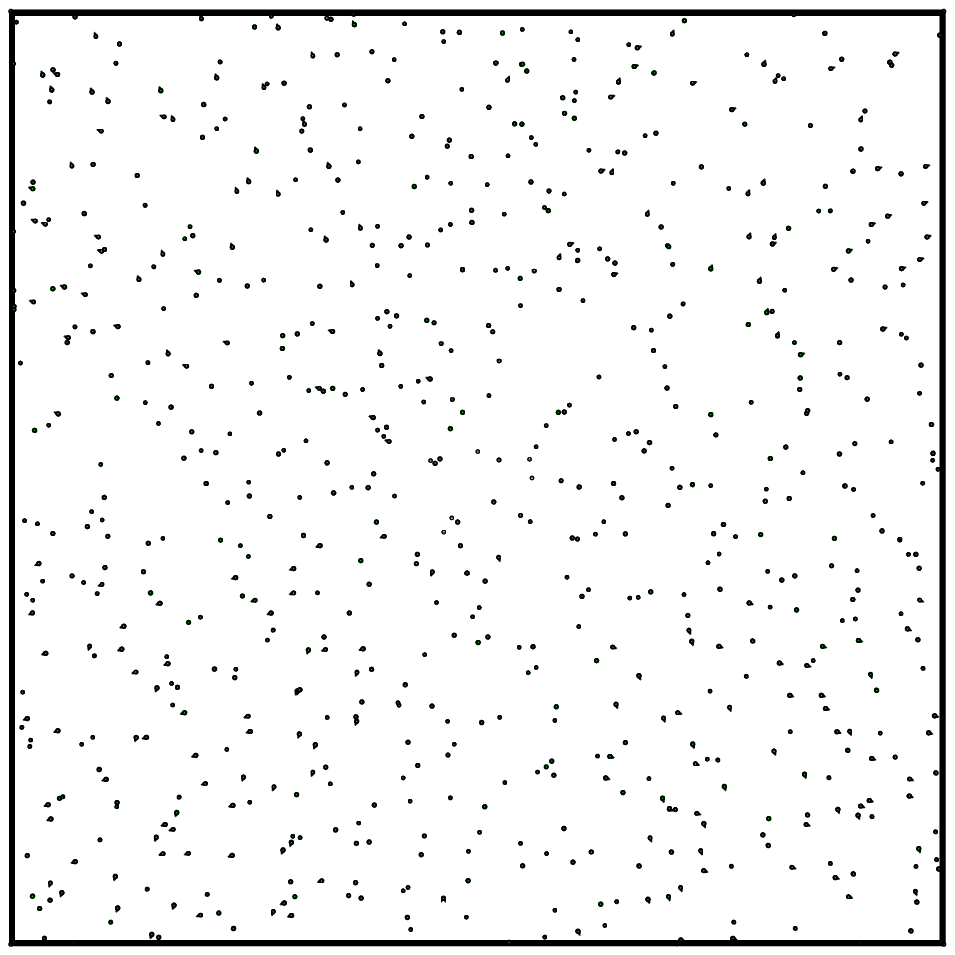}\bigskip}
 {\includegraphics[width=3.5cm,angle=-90]{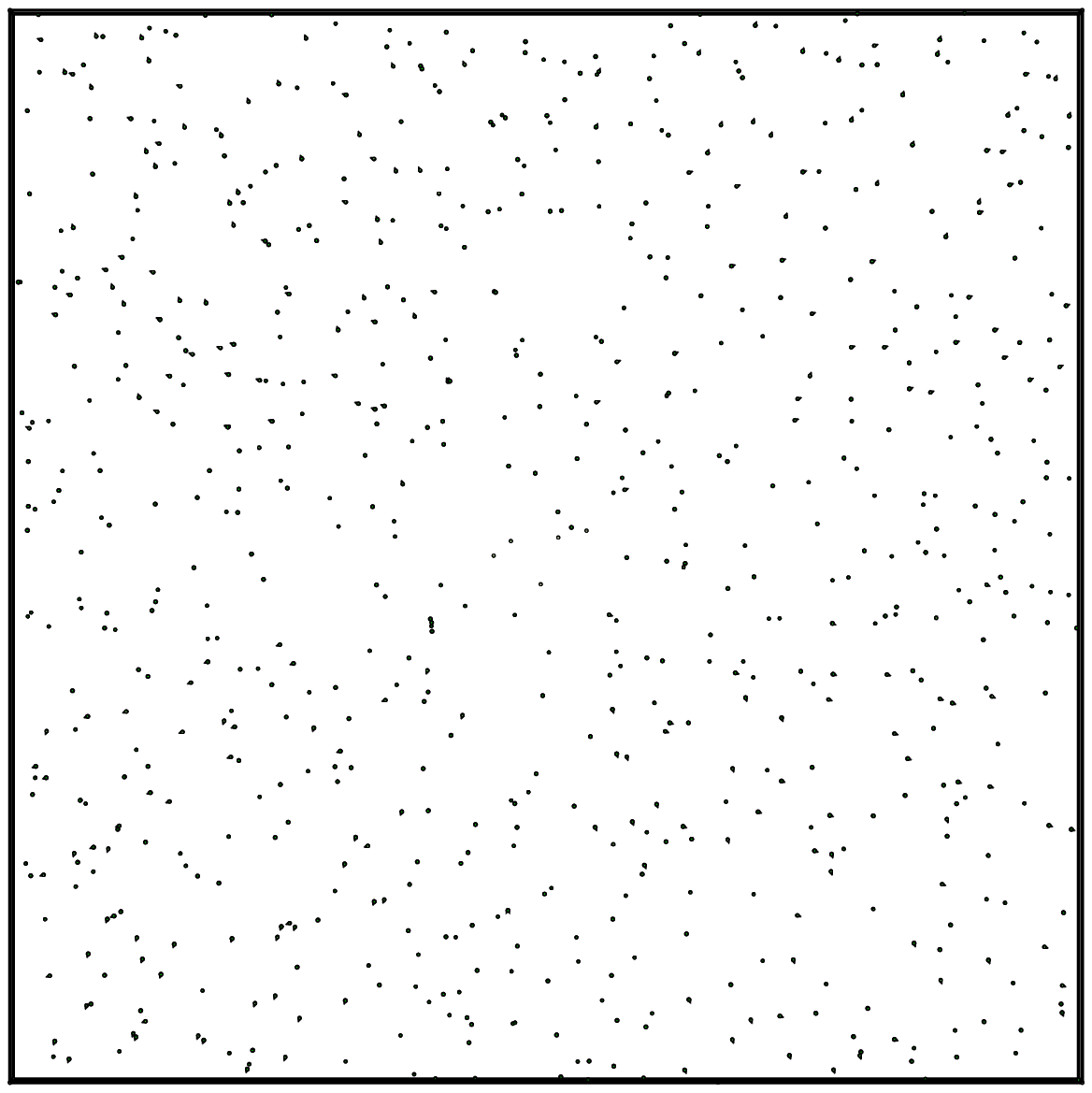}\bigskip}
 \end{minipage}
 \begin{minipage}[!t]{3\dimen0}
 \begin{center}
 {\includegraphics[width=3.5cm,angle=-90]{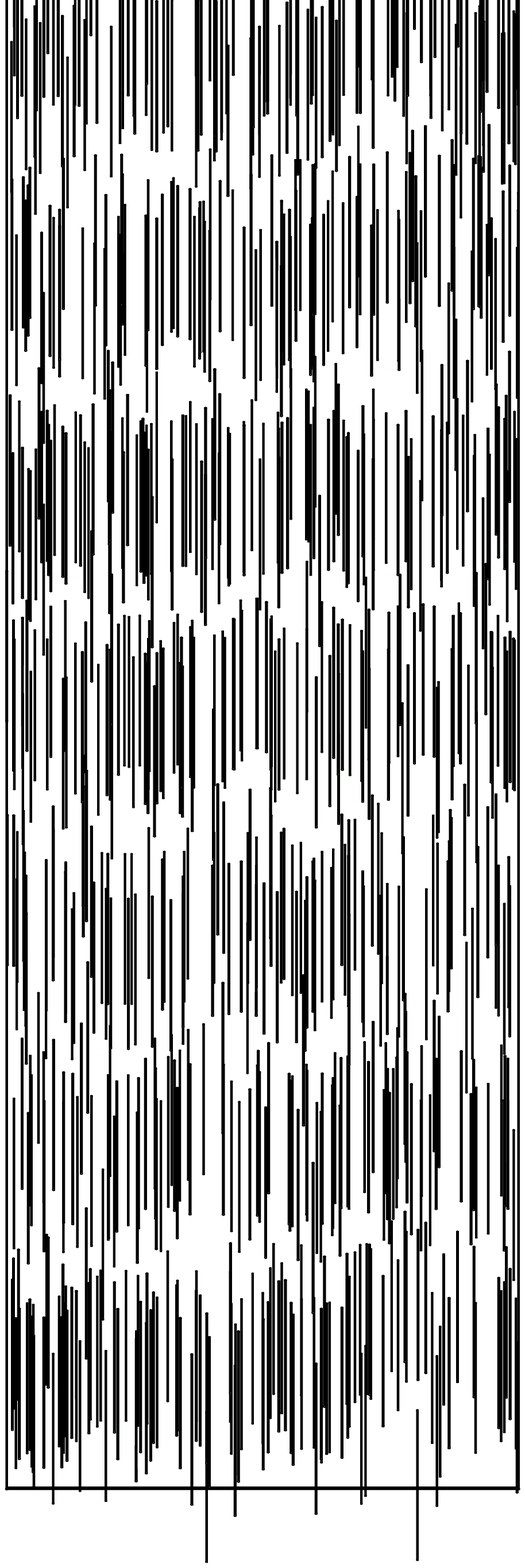}\bigskip}
 {\includegraphics[width=3.5cm,angle=-90]{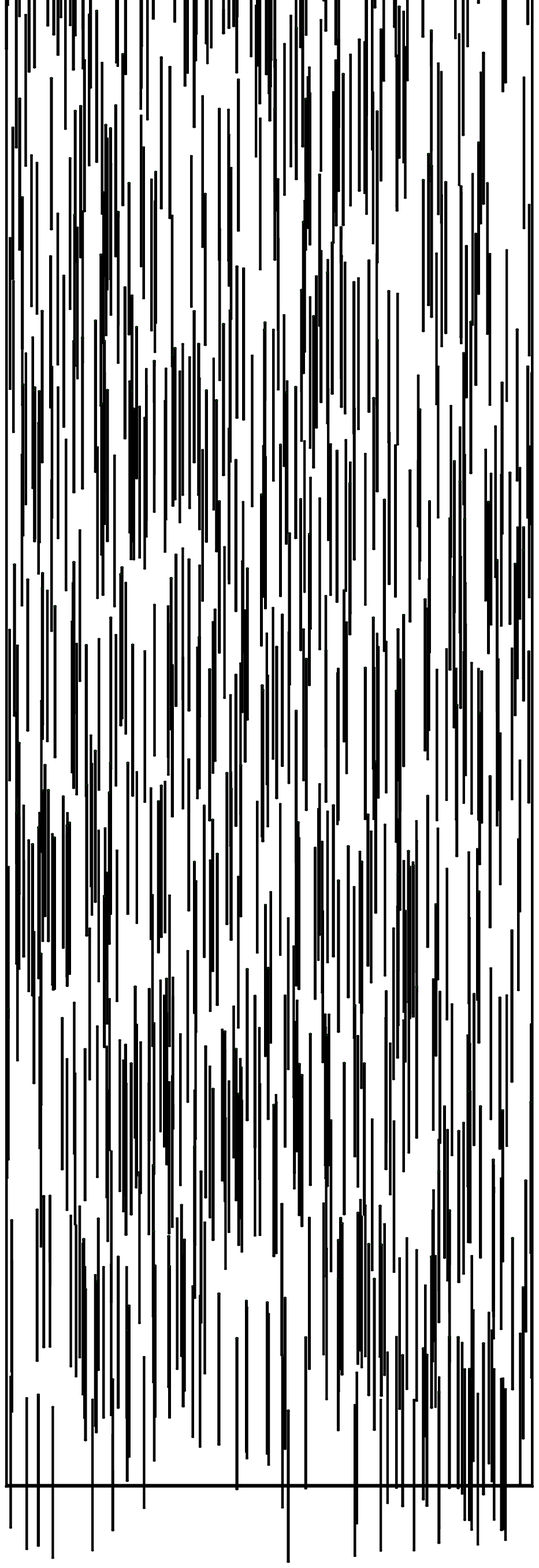}\bigskip}
 {\includegraphics[width=3.5cm,angle=-90]{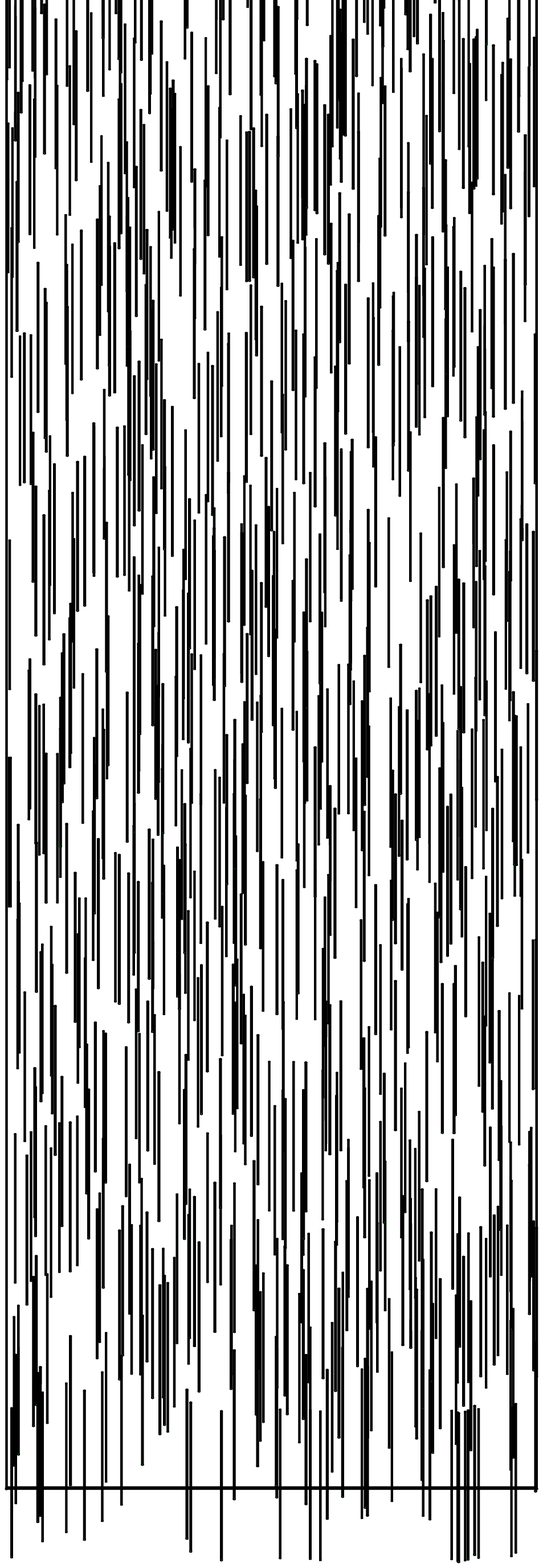}\bigskip}
 {\includegraphics[width=3.5cm,angle=-90]{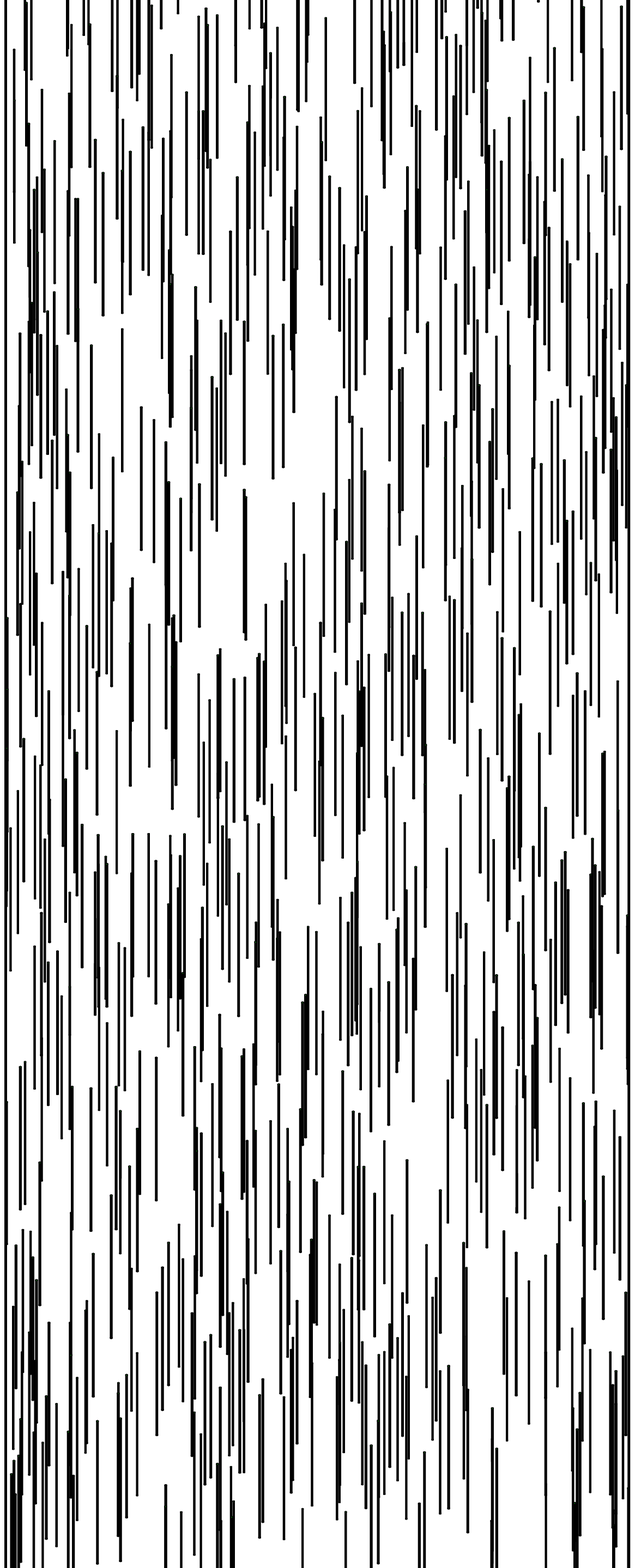}\bigskip}
 \end{center}
 \end{minipage}
\caption{Snapshots of equilibrated configurations
for (from top to bottom) $P^*=2.50 \ (\langle\rho^*=0.23\rangle)$,
$P^*=2.30 \ (\langle\rho^*=0.44\rangle)$, $P^*=2.00 \
(\langle\rho^*=0.47\rangle)$,  and $P^*=0.50 \
(\langle\rho^*=0.49\rangle)$. Left panels: projections of the
centers of mass of the spherocylinders onto the $x$-$y$ plane. Right
panels: configurations along the $z$ axis. }\label{fig:2}
\end{figure*}

We verified that, during the simulation, the system spontaneously
transforms into a smectic phase, as can be seen through the
snapshots of equilibrated configurations that are shown in
Fig.~\ref{fig:2}. Such snapshots were taken for densities across
the transition threshold. Traces of smectic order can be hardly
detected for $P^*=2.00$, a pressure corresponding to an average
density equal to $0.44$. The ordering of the fluid is already
apparent just beyond the transition threshold for $P^*=2.30$ and
$\rho^*\simeq 0.47$. Fully developed arrangements of several
well-separated smectic layers distributed along the $z$ direction
are clearly visible for $P^*=2.50$ and $\rho^*\simeq 0.49$.
Instead, the transversal arrangements of the spherocylinders along
the $x$-$y$ plane look disordered in all states investigated
(compare left and right panels in Fig.~\ref{fig:2}).

\begin{figure*}[t]
\begin{center}
\includegraphics[width=9.0cm,angle=-90]{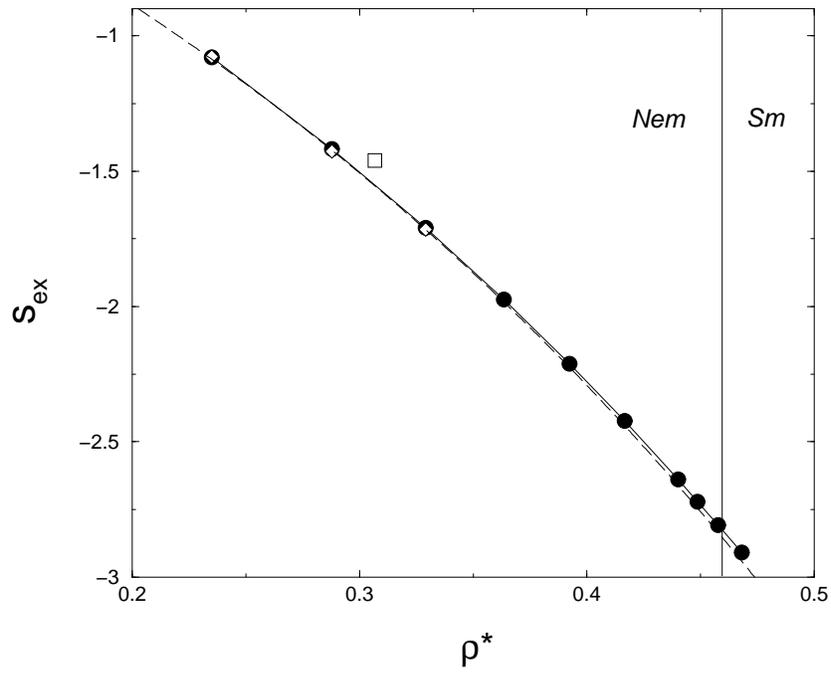}
\caption{Excess entropy plotted as a function of the reduced
density. Line with solid circles: this work with $N = 1500$;
diamonds: Widom insertion-method estimates. Square: multistage
Widom test~\protect\cite{koda}.
Dashed line: excess entropy evaluated by integration of the
five-term virial fit of the simulation data of
Ref.\protect\cite{slf}. The vertical line indicates the nematic-smectic
transition
threshold according to Ref.~\protect\cite{slf}
and~\protect\cite{verfre}.}\label{fig:3}
\end{center}
\end{figure*}

We show in Fig.~\ref{fig:3} the excess entropy of the fluid
plotted as a function of the reduced density. The data were
obtained using both the thermodynamic integration of
Eq.~(\ref{s:free_ex}) as well as Widom's ghost-particle method
implemented at moderately low densities. We also show the datum
recently reported by Koda and Ikeda~\cite{koda} which was obtained
using a multistage Widom test, based on the gradual insertion of a
ghost particle with a variable shape. The modest discrepancy
between this finding and the present results is likely due to the
multistage
method, that is known to work better at high densities. On the
other hand, our results for the excess entropy are closely
interpolated by the expression obtained upon integrating the
five-term virial fit of the EoS reported in~\cite{slf}.

\begin{figure*}[t]
\flushleft
\includegraphics[width=6.60cm,angle=-90]{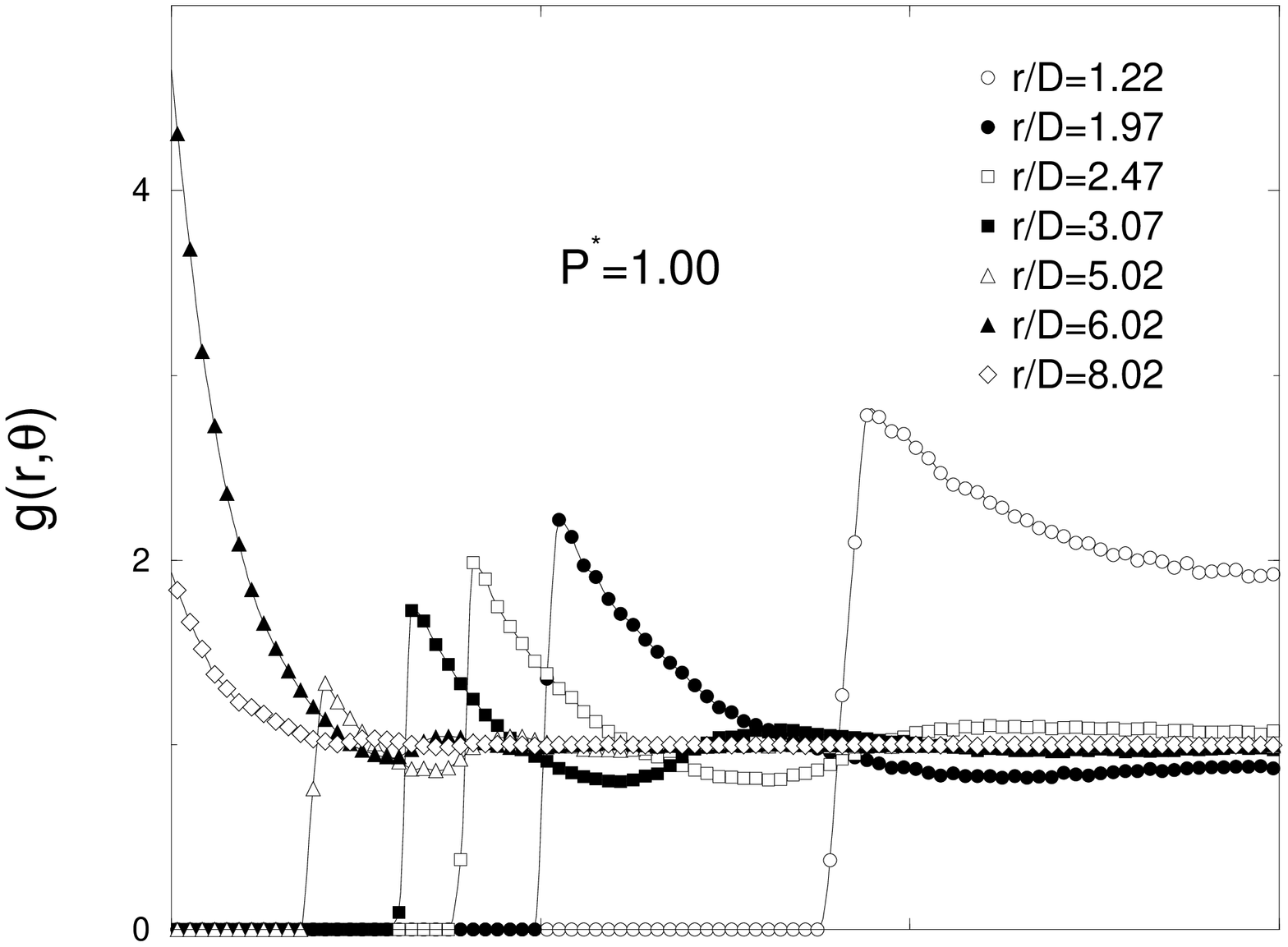}
\includegraphics[width=6.55cm,angle=-90]{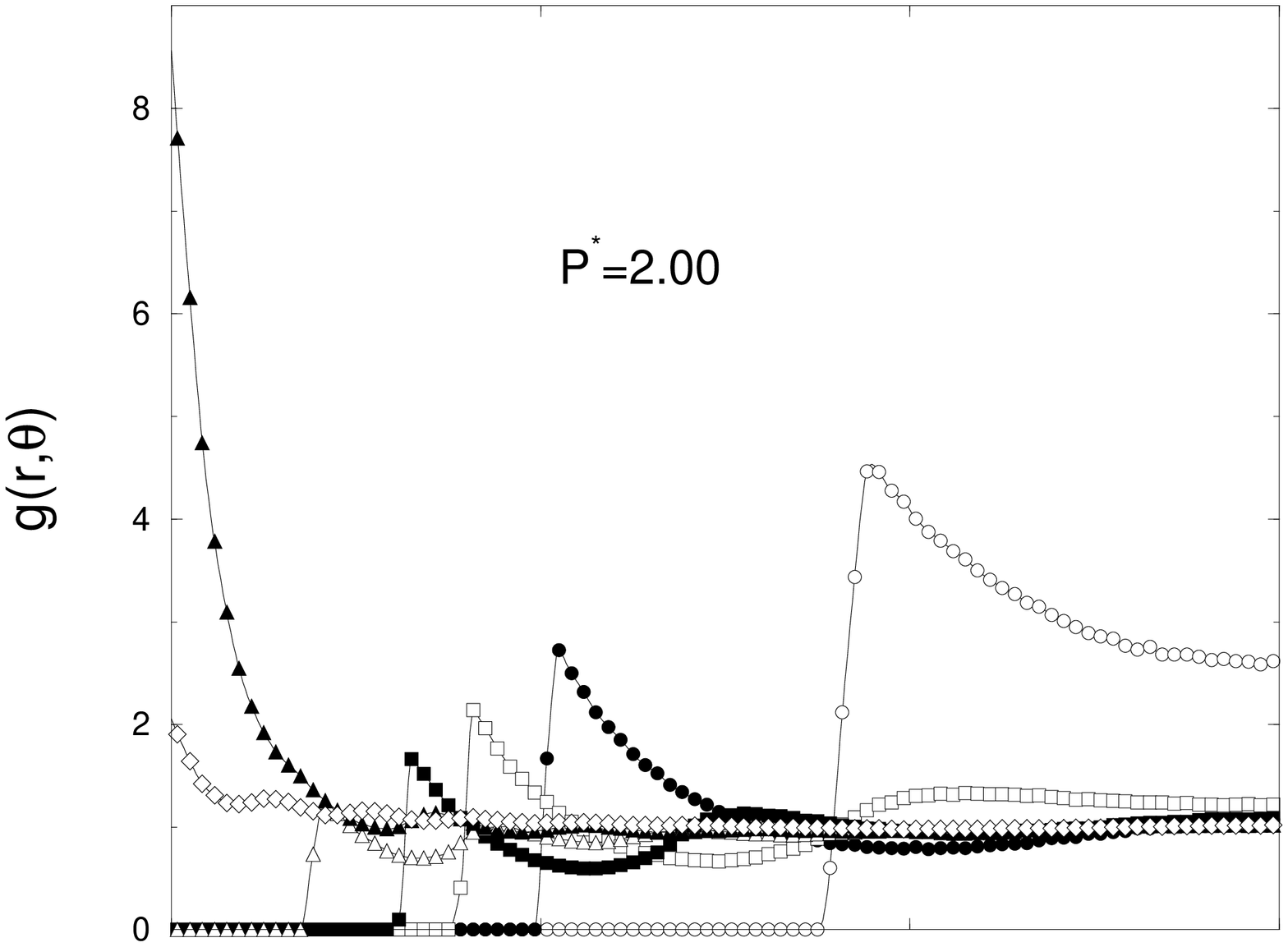}
\includegraphics[width=8.35cm,angle=-90]{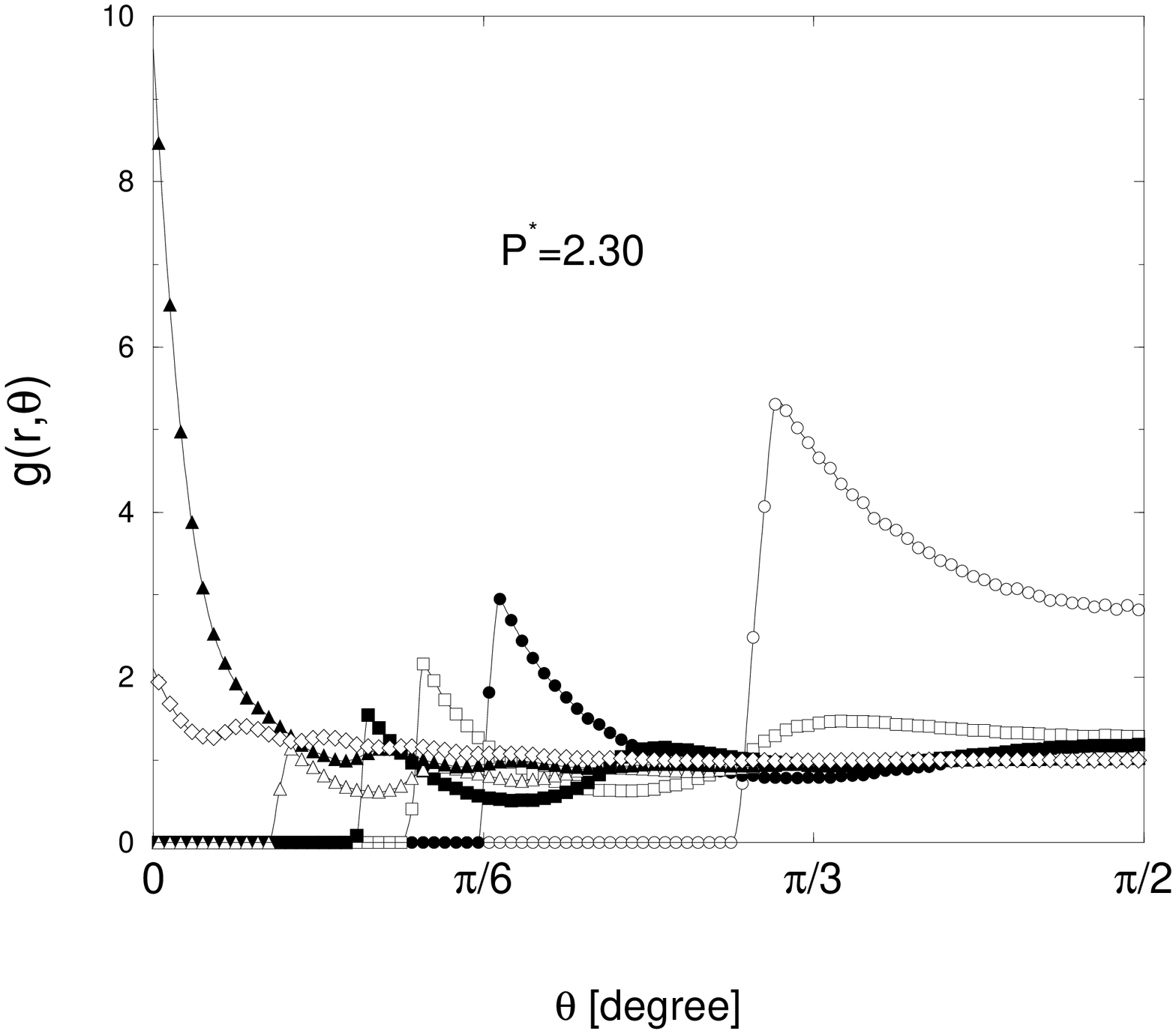}
\begin{center}
\caption{Pair distribution function
$g(r,\vartheta)$ plotted as a function of $\vartheta$ for
different distances $r/D$ and for increasing
pressures.}\label{fig:4}
\end{center}
\end{figure*}

The PDF of the fluid, $g(r,\vartheta)$, was plotted in
Fig.~\ref{fig:4} as a function of $\vartheta$ for a set of
interparticle distances at increasing pressures across the
transition point ($P^*\simeq 2.2$). We note, for separations $r <
L+D$, the existence of a forbidden range around $\vartheta = 0$
and $\pi$ that is due to the overlap of two spherocylinders. This
correlation gap decreases with increasing intermolecular
separations. The maximum attained by $g(r,\vartheta)$ corresponds
to the hard-core contact between spherocylinders and its height
consistently decreases with increasing interparticle distances.
For $r \ge L+D$, the entire angular range between $0$ and $\pi$
can be sampled by a second spherocylinder. Of course, there is a
blow up of the contact value of the PDF for $\vartheta=0,\pi$,
which progressively decreases for increasing distances. An
increase of the pressure enhances the overall structure of the
PDF; however, no specific signature of the nematic-smectic
transition can be detected in the resulting modification of the
PDF.

\begin{figure*}
\begin{center}
\includegraphics[width=9.0cm,angle=-90]{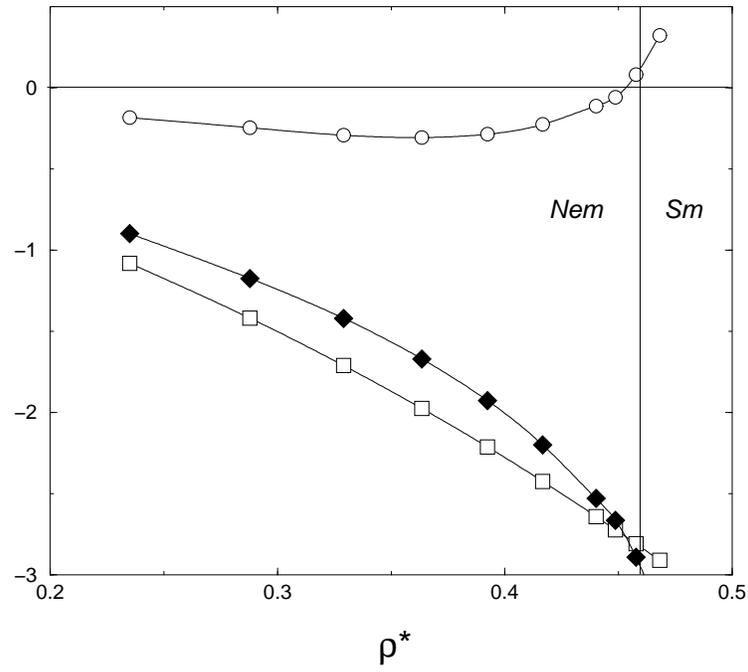}
\caption{ Residual multiparticle entropy (circles) resolved into
the excess (squares) and pair (solid diamonds) contributions.
Lines are smooth interpolations of the simulation data.
}\label{fig:5}
\end{center}
\end{figure*}

Upon plugging the PDF into Eq.~(\ref{eq:s2}), we obtained the pair
entropy that was plotted together with the excess entropy in
Fig.~\ref{fig:5} as a function of the density. The RMPE exhibits a
change of sign from negative to positive values for $\rho^*\simeq
0.453$. This threshold practically coincides with the currently
accepted nematic-smectic transition density
($\rho^*=0.46$)~\cite{slf,verfre}. The simulations carried out for
$500$ particles enclosed in a cubic box show a similar behavior of
the RMPE whose morphology is not substantially modified with
respect to that of $1500$ spherocylinders in an orthorombic box.
This finding is noteworthy in that it demonstrates the high
sensitivity of the RMPE to the structural changes occurring in the
fluid, irrespectively of the difficulty to accommodate a well
resolved smectic layering in a relatively small cubic box.

\begin{figure*}
\begin{center}
\includegraphics[width=9.0cm,angle=-90]{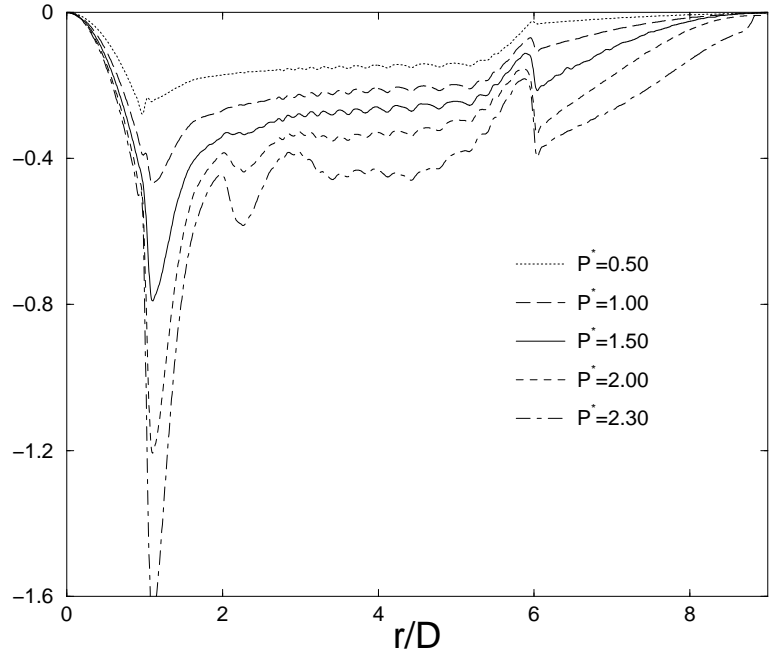}
\caption{Integrand function appearing in Eq.~(\ref{eq:s2}), after
integration over $\vartheta$, plotted for several
pressures.}\label{fig:6}
\end{center}
\end{figure*}

Figure~\ref{fig:6} shows the quantity that, upon integration,
yields the pair entropy in Eq.~(\ref{eq:s2}). For very short
interparticle distances the $\vartheta$ range which can be sampled
by two neighboring particles is rather limited. As a result,
angular correlations are strong and give rise to the deep well in
the integrand function. A steady modulated increase follows up to
$r=L+D$ where a negative jump witnesses the onset of new strong
correlations between spherocylinders lying on top or below the
central reference particle. This effect is the integrated
counterpart of the behavior that was already discussed for $g(r,
\vartheta)$ in Fig.~\ref{fig:4}.

\begin{figure*}
\begin{center}
\includegraphics[width=9.0cm,angle=-90]{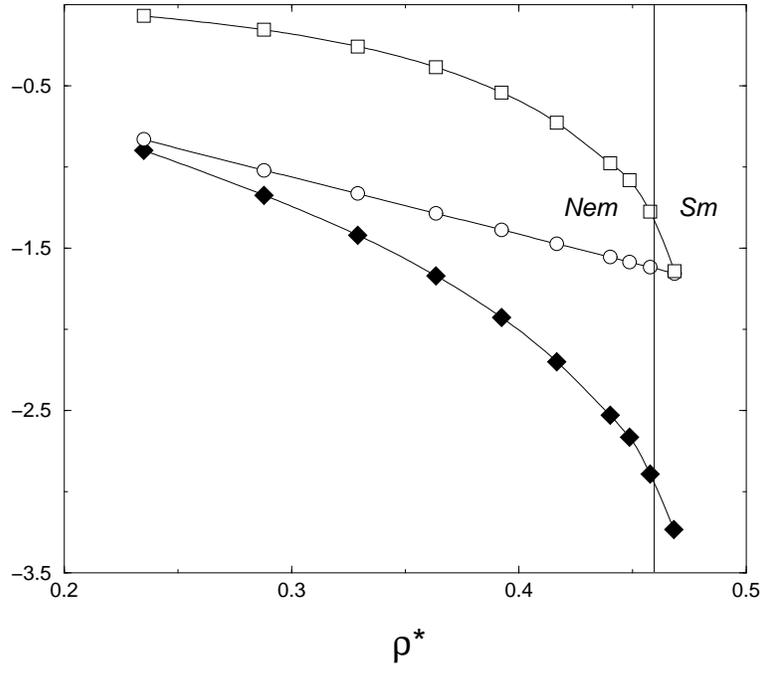}
\caption{Pair entropy (diamonds) resolved into the excluded-volume
(circles) and correlation (squares) contributions (see
Eq.~(\ref{s:s2})). Lines are smooth interpolations of the
simulation data.}\label{fig:7}
\end{center}
\end{figure*}

The pair entropy was finally resolved into an excluded-volume and
a correlation term in Fig.~\ref{fig:7} (see Eq.~\ref{s:s2}). It is
clear that the second-order virial term cannot account by itself
for the crossover between $s_{\rm ex}$ and $s_2$ that is
indicative of the smectic ordering of the nematic fluid.

\begin{figure*}
\begin{center}
\includegraphics[width=9.0cm,angle=-90]{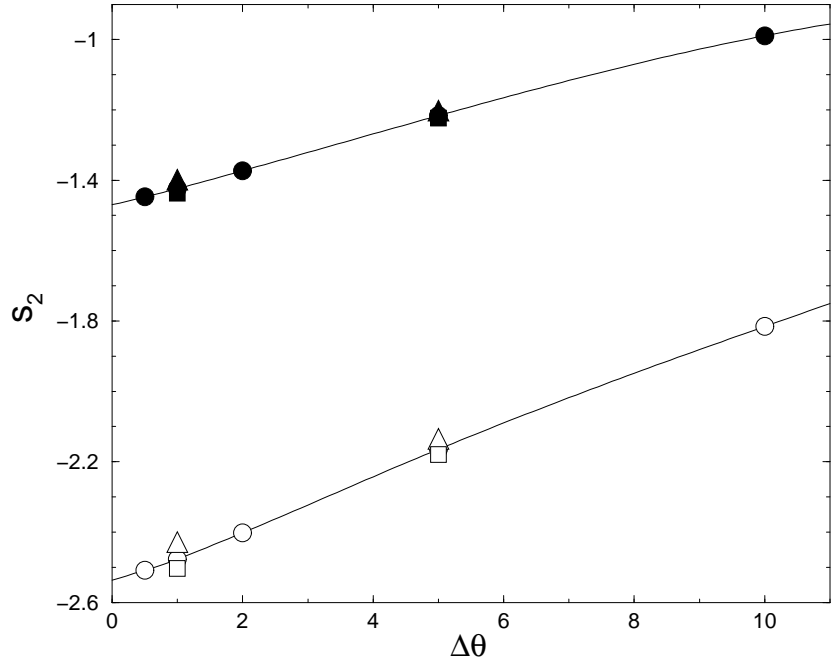}
\caption{The pair entropy $s_2$ plotted as a function of the grid
mesh parameter $\Delta\vartheta$ and for several choices of
$\Delta r$, for $P^*=1.00$ (solid symbols) and $P^*=2.00$ (open symbols). 
Triangles, $\Delta r=0.10D$; circles, $\Delta
r=0.05D$; squares $\Delta r=0.02D$. }\label{fig:8}
\end{center}
\end{figure*}

As for the numerical reliability of the current results, we note
that the statistics cumulated on $g(r,\vartheta)$ was such that a
smooth integration over the sampled points yielded stable values
for $s_2$, the dispersion being always lower than a percent of the
average value. However, we observed a moderate sensitivity of the
angle-dependent quantities in Eq.~(\ref{eq:s2}) on the resolution
of the angular width $\Delta \vartheta$. In order to gain a better
insight into the numerical accuracy of the calculations, we
performed a series of test runs for two different pressures
($P^*=1.0$ and $2.0$) with several $r$ and $\vartheta$ grid
meshes. The tests were reported in Fig.~\ref{fig:8} where the
higher sensitivity of $s_2$ on the angular grid size~---~as compared
with the radial one~---~is quite manifest. We estimated, in the
limit of $\Delta r, \Delta \vartheta \to 0$, an asymptotic shift
of $s_2$ toward lower values not larger than $2\%$. As a result,
we expect a comparable shift of the transition threshold signalled
by the RMPE to lower densities.

\section{Concluding remarks}

In this paper we have analyzed the residual multiparticle entropy
(RMPE) of parallel hard spherocylinders with aspect ratio $L/D=5$
across the nematic-smectic
transition. The correspondence between the intrinsic ordering
threshold detected through the vanishing of the RMPE
($\rho^*\simeq 0.453$) and the independently ascertained
phase-transition density ($\rho^*=0.46$) is quantitative. It also
turned out that the indication of the zero-RMPE criterion is not
significantly affected by the size and shape of the simulation
box, even when the smectic layering of the fluid cannot be easily
accommodated in the sample as is the case of a small cubic box.
This finding further corroborates the sensitivity of the RMPE to
the ordering of the fluid on a local scale.

\newpage

\end{document}